\documentclass[aps,prl,twocolumn,floatfix,superscriptaddress]{revtex4}

\usepackage{amsmath,amssymb,amsfonts}
\usepackage{bm}
\usepackage{graphicx}
\usepackage{times}
\usepackage{color}
\usepackage[colorlinks,bookmarks=false,citecolor=blue,linkcolor=red,urlcolor=blue]{hyperref}

\newcommand{\fref}[1]{Fig.~\ref{#1}}
\newcommand{\tref}[1]{Table~\ref{#1}}
\newcommand{\figcite}[1]{{\protect \cite{#1}}}

\newcommand{\tj}{$t{-}J$ }

\newcommand{\JJC}{ \exp\left( \sum_{i,j=1,N}{v^c_{|i-j|} n_i n_j}\right)}
\newcommand{\JJS}{ \exp\left( \sum_{i,j=1,N}{v^S_{|i-j|} S^z_i S^z_j}\right)}

\newcommand{\dz}{$d_{3z^2-r^2}$ }

\newcommand{\normwidth}{0.8\columnwidth}
\newcommand{\smallwidth}{0.66\columnwidth}

\newcommand{\miniwidth}{0.49\columnwidth}

\begin{document}

\title{Orbital currents in extended Hubbard models of high-T$_c$ cuprates}
\author{C\'edric Weber}
\affiliation{Dpt. of Physics and Astronomy, Rutgers University, NJ-08854 Piscataway, USA}
\author{Andreas L\"auchli}
\affiliation{Institut Romand de Recherche Num\'erique en Physique
  des Mat\'eriaux (IRRMA), CH-1015 Lausanne, Switzerland}
\author{Fr\'ed\'eric Mila}
\affiliation{Institute of Theoretical Physics, EPFL, CH-1015
  Lausanne, Switzerland}
\author{T. Giamarchi}
\affiliation{DPMC-MaNEP, University of Geneva, 24 Quai Ernest-Ansermet CH-1211 Geneva, Switzerland }

\begin{abstract}
Motivated by the recent report of broken time-reversal symmetry and zero momentum magnetic scattering
in underdoped cuprates, we investigate under which circumstances orbital currents circulating inside
a unit cell might be stabilized in extended Hubbard models that explicitly include oxygen orbitals.
Using Gutzwiller projected variational wave functions that treat on an equal footing all instabilities,
we show that orbital currents indeed develop on finite clusters, and that they are stabilized in
the thermodynamic limit if additional interactions, e.g. strong hybridization with apical oxygens,
are included in the model.
\end{abstract}
\maketitle

Despite intensive efforts in the last twenty years, the physics of high Tc superconductors remains largely mysterious
\cite{bonn_hightc_review}. This is especially true of the pseudogap phase of underdoped cuprates, for which various explanations have been put forward ranging from preformed superconducting pairs \cite{kotliar_liu_dwave_slavebosons}
to the existence of orbital currents with \cite{chakravarty_ddw_pseudogap} or without \cite{varma_first_time_mf_theta2,varma_pseudogap_theory} broken translational symmetry. The latter case has received considerable attention due on one hand to recent neutron experiments \cite{fauque_neutrons_currents} indicating the presence of magnetic
moments, compatible with the translation invariant pattern of currents predicted by Varma \cite{varma_first_time_mf_theta2}, and on the other hand to Kerr effect measurements
\cite{xia_polarkerr_pseudogap} showing evidence of time-reversal symmetry breaking.

Current (flux) phases have been first proposed for the single-band Hubbard model \cite{affleck_marston,kotliar_liu_dwave_slavebosons,laderersuperconductivity_flux_ref,zhang_flux_ref}, but have
been found unstable by slave bosons and numerical calculations. Interestingly, orbital currents were also found to be relevant in two-band system \cite{coleman_flux_2band,coleman_flux_2band2}. In ladder models, where the existence of such phases can be checked in a more controlled way, it was found that somewhat special interactions, more complex than local ones, are needed to stabilize them \cite{orignac_2chain_long,schollwock_ladder_strong_fluxphases}. The resulting phases break the translational symmetry of the lattice, leading to a staggered flux pattern. Similar staggered patterns were advocated as a potential explanation of the pseudogap phase \cite{chakravarty_ddw_pseudogap} (DDW phase). 

To stabilize flux phases that do not break the translational symmetry,
it seems worthwhile to go beyond the single band Hubbard model and to consider the so-called three-band Hubbard model in which oxygen orbitals are explicitly taken into account. While this model has been tested for superconducting instabilities early on, pointing to some differences with the single band model \cite{sudbo_cuo}, the investigation of
its flux phase instabilities has only started quite recently. The one-dimensional (ladder) case has first been investigated at half filling, when the system is essentially an insulator \cite{lee_marston_CuO}. More recently, a low energy analysis for the three band model on a ladder has been performed, and showed that in a certain range of doping flux phases were indeed stabilized \cite{chudzinski_ladder_rapid}. These phases exhibit a $2k_F$ staggered order parameter, quite natural in one-dimension, for the currents, but with a symmetry different from that of the DDW.
In two dimensions, a mean-field analysis \cite{varma_first_time_mf_theta2} of the three band model has suggested the existence of translationally invariant current patterns when the Cu-O nearest-neighbor repulsion is strong enough. This result has not been
confirmed by exact diagonalizations \cite{greiter_exactdiag_currents,greiter_exactdiag_currents2}
on small clusters of an effective \tj model. However
the mapping of the three-band Hubbard model on this \tj model can only be justified in the limit
of very large oxygen on-site energy, a situation
not realized in the cuprates \cite{hybertsen_DFT_parameters_cuprates}. Moreover, because of the three atoms per unit cell, the exact diagonalizations have been performed for clusters with only a few unit cells, and the relevance
of the results for the thermodynamic limit is far from guaranteed, in particular because the filling that was considered on this small cluster $x=12.5\%$ is leading to a polarized ground-state with finite momentum,
which is not representative of the physics on large scales.  Therefore, further investigations of the three-band Hubbard model are clearly called for.

\begin{figure}
  \begin{center}
    \includegraphics[width=\smallwidth]{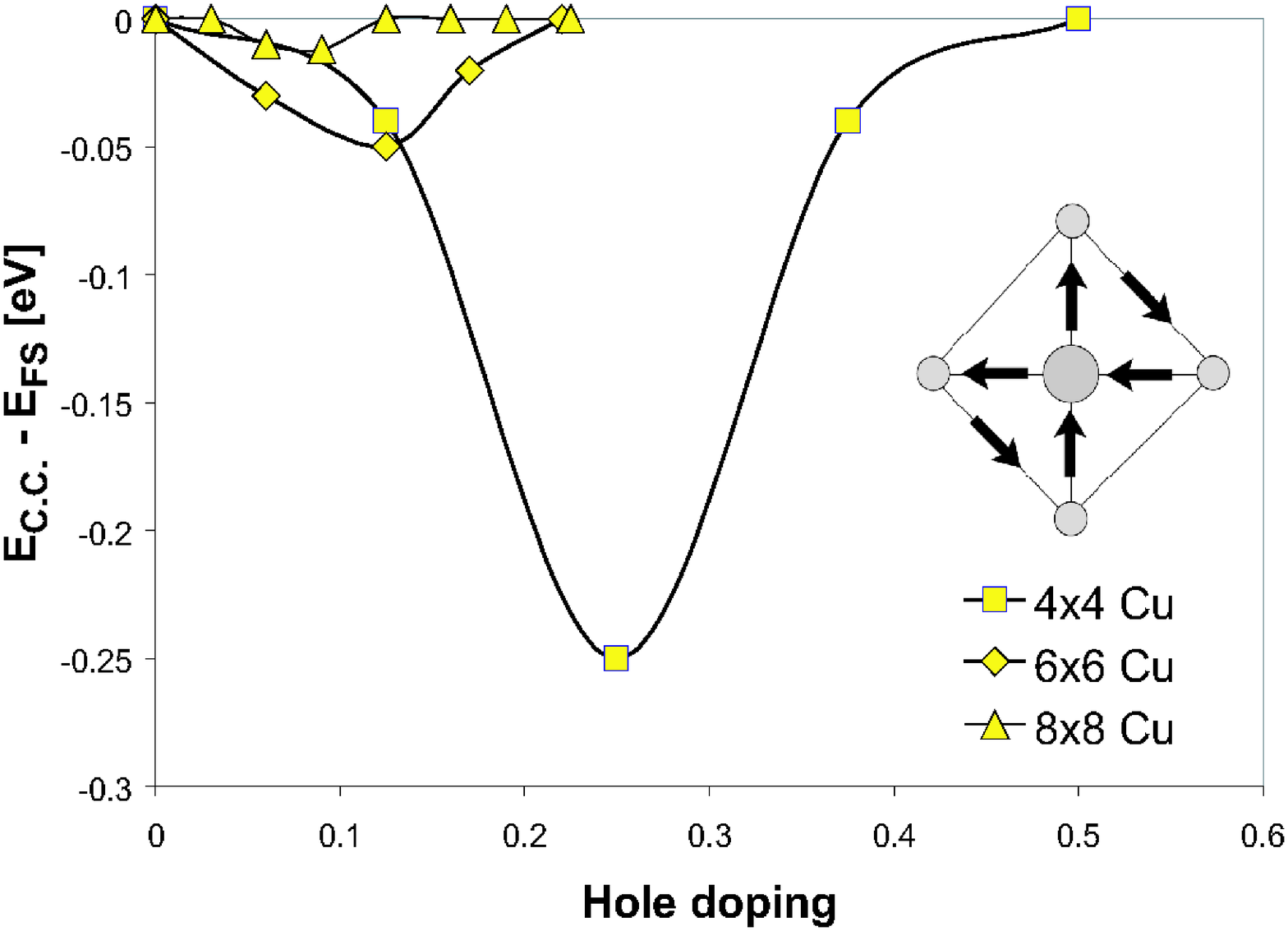}
    \includegraphics[width=\smallwidth]{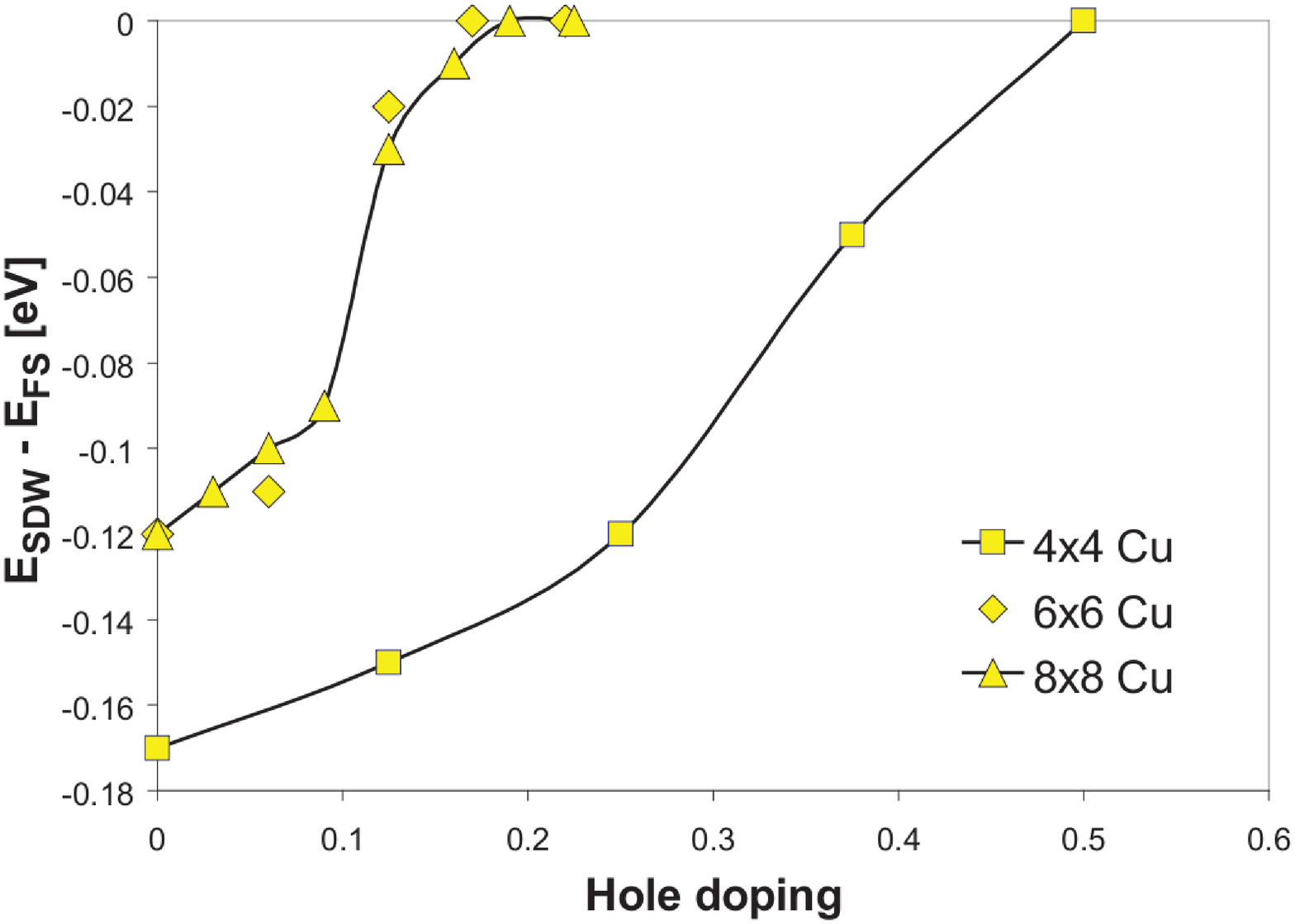}
    \caption{Energy of the best orbital current wavefunction (top) and of the best spin density wave wavefunction (bottom). The reference was taken as the projected Fermi sea (full Jastrow projector). Inset: the symmetry of the best current phase found ($\Theta_2$ pattern in the notations of \figcite{varma_first_time_mf_theta2}).
Although the orbital current phase is stable on an intermediate size $16$ copper lattice, the energy gain is
strongly reduced when the size increases. To the accuracy of the calculation,
we did not find any energy optimization on a $100$ copper lattice for the orbital current instability.} \label{fig:energies}
  \end{center}
\end{figure}

In this paper we perform a variational Monte Carlo (VMC) investigation of the three band Hubbard model based on a Gutzwiller projected wave function that allows for the possibility of orbital currents. This provides an unbiased method, free from numerical limitations even for large system sizes, for which current instabilities are treated on an equal footing with other instabilities.
We find that on intermediate system sizes, a current flux phase circulating between copper and oxygens is stabilized.
This phase has the same symmetry ($\Theta_2$, see \fref{fig:energies}) as the phase found in the mean field solution \cite{varma_first_time_mf_theta2}.
Other symmetries or phases that break the translational symmetry are much higher in energy.
However, as the system size gets larger, the energy gain decreases strongly, making it unclear whether such a phase would survive in the thermodynamic limit. We propose modification of the Hamiltonian that takes into account apical oxygens or three-body terms and which strongly stabilizes such current patterns.

The three band model is defined by the Hamiltonian:
\begin{multline}
\label{eq:3band_hub}
H=\sum_{\langle i,j \rangle \sigma}{\left(t_{i,j} d^\dagger_{i\sigma}p_{j\sigma} + c.c. \right)} +
  \sum_{\langle\langle i,j \rangle\rangle \sigma}{\left(t_{i,j} p^\dagger_{i\sigma}p_{j\sigma} + c.c. \right)} \\
 + U_p \sum_{p}{ \hat n_{p\uparrow}} \hat n_{p\downarrow} + U_d \sum_{d}{ \hat n_{d\uparrow} \hat n_{d\downarrow}} + \\
\Delta_p \sum_{p,\sigma}{\hat n_{p\sigma}} + V_{dp} \sum_{d,p}{\hat n_d \hat n_p}
\end{multline}
where $t_{i,j}$ stand for the hopping matrix elements between nearest-neighbor Cu-O pairs ($\langle i,j \rangle $),
of magnitude $t_{dp}$, and between next-nearest neighbor O-O pairs ($\langle\langle i,j \rangle\rangle$), of magnitude $t_{pp}$, while $\Delta_p$, $U_d$, $U_p$ and $V_{dp}$ denote the atomic energy of the O-$p$ orbitals, the
on-site repulsions in the Cu-$d$ and O-$p$ orbitals, and the
nearest-neighbor repulsion between Cu-$d$ and O-$p$ orbitals. A realistic set of parameters
found by LDA calculations \cite{emery_3bands_model,hybertsen_DFT_parameters_cuprates,mila_parameters_cuprates} is
$U_d=10.5$ eV and $U_p=4$ eV, $|t_{dp}|=1.3$ eV and $|t_{pp}|=0.65$ eV, $\Delta_p=3.5$ eV, and $V_{dp}=1.2$ eV, and,
unless specified otherwise, these values are used in the following. Note that we work in hole notations.
The phase factor that comes from the hybridization of the $p-d$
orbitals \footnote{In hole notations the bonding orbitals enter the Hamiltonian with a positive transfer integral sign, and the anti-bonding orbitals with a negative sign. In electron notations, the signs are reversed.} gives to the $t_{ij}$ a non homogeneous sign.

Around each copper atom, there is one Cu-O-Cu plaquette with three minus signs. Moreover, the product of the hopping signs around all the Cu-O-Cu plaquette is $-1$ in hole notations. Interestingly, a simple gauge transformation involving a double copper unit-cell leads to $t_{ij}=-|t_{ij}|$ \footnote{This is used within our exact diagonalization study to preserve the rotation symmetries and hence to reduce the size of the Hilbert space sectors.}.

This model has already been investigated with variational Monte Carlo \cite{yanagisawa_vmc_3band,oguri_variational_3band},
but the wave functions used did not allow for current instabilities.
By contrast, the wave function we consider throughout this work is constructed from the ground state of the Hofstadter-like
mean-field hamiltonian:
\begin{equation}
\label{eq:fullmf}
H_{MF} = \sum\limits_{(i,j)}{t_{ij}|\chi_{ij}|e^{\imath \theta_{ij}} c^\dagger_{i\sigma}c_{j\sigma}}
+ \Delta^{var}_p \sum_{p\sigma}{\hat n_{p\sigma}} + \sum_{i}{\bold{h_i}\cdot\bold{S_i}}
\end{equation}
where $\chi_{ij}>0$,$\theta_{ij}$, $\Delta_p^{var}$ and $\bold{h_i}$ are real variational parameters.
In this expression, the sum runs over nearest and next-nearest neighbor pairs, and $c$ stands for $p$ or $d$
orbitals depending on the site.
The local magnetic field
$\bold h_i$ allows to consider antiferromagnetism.
% $\Delta_p^{var}$ is
%the renormalized energy difference between the $d$ and $p$ atomic levels.
The 16 variables $\chi_{ij}$ and $\theta_{ij}$ are independent within one copper unit-cell.
To deal with staggered order we multiply a subset of the variational parameters by -1.
When $\theta_{ij}\ne 0$, the order parameters are associated with an external flux which leads to the circulation of the holes. In this case the Green function is complex:
$\langle c^\dagger_i c_j \rangle = |\langle c^\dagger_i c_j \rangle|\ \exp(\mathrm{i} \phi_{ij}) $
We treat the correlations with a spin and charge Jastrow factor:
\begin{equation}
 \label{eq:real_jastrow}
   \mathcal J = \JJC \JJS
\end{equation}
where all $v_{|i-j|}^c$ and $v_{|i-j|}^S$ are considered as free
variational parameters.

We are mainly interested in the charge current observable. The conservation of the density:
\begin{equation}
\label{eq:cons_current}
  \frac{\delta n_i}{\delta t} = 0 = \frac{\hbar e}{c} \left[H , n_i \right] = \sum\limits_{\langle i,j \rangle}{J_{i,j}}
\end{equation}
leads to the current operator \footnote{The derivative of the Hamiltonian
with respect to the vector potential $\mathcal{A}_{ij}$ leads to the same definition for the current.} on a link:
$J_{ij} = \sum\limits_{\sigma}{ t_{ij} |\langle c^\dagger_i c_j \rangle| \sin\left( \phi_{ij}  \right)}$.
However the current conservation law is not satisfied by our variational wave function
since it is not an exact eigenstate of (\ref{eq:3band_hub}). The \emph{mean-field current},
defined as $J^{MF}_{ij} := \sum\limits_{\sigma}{ t_{ij}|\chi_{ij}||\langle c^\dagger_i c_j \rangle| \sin \left(\theta_{ij}+\phi_{ij} \right)}$
is a conserved quantity, but in general $J$ and $J_{MF}$ need not be oriented in the same way.
In particular we find that the O-O currents have different signs for $J$ and $J_{MF}$. The phase
$e^{\imath\theta_{ij}}$ thus gives the direction of $J^{MF}$, but \emph{not} the direction of $J$ in general, which has to be computed explicitly.

In order to overcome this difficulty and to impose the current conservation at each vertex of the lattice, we apply on
the wavefunction an additional complex Jastrow factor \cite{Trivedi_complex_Jastrow}:
\begin{equation}
\label{eq:complex_jastrow}
   \mathcal J_c =  \exp\left( \sum_{i=1,N}{ i \alpha_{i} n_i }\right)
\end{equation}
The kinetic energy of the extended wavefunction is $\langle T \rangle = \sum\limits_{\langle i,j,\sigma \rangle}{t_{ij}|\langle c^\dagger_{i\sigma} c_{j\sigma} \rangle| \cos\left(\theta_{ij} +\alpha_j-\alpha_i\right) }$.
We emphasize that the variables $\alpha_i$ are in general not able to cancel the flux $\theta_{ij}$.
The minimization of the variational parameters introduced in (\ref{eq:fullmf}) and in (\ref{eq:real_jastrow})
is performed using a stochastic minimization procedure
\cite{umrigar_mcv_minimis,sorella_optimization_VMC} in which
the parameters of the uncorrelated part of the wavefunction and
the Jastrow parameters are minimized at the same time. This method allows one to deal with
a large number of parameters since the gradients are calculated all at the same
time during a simulation. The new parameters are then calculated using the obtained
gradients, and the procedure is iterated until convergence. At each step, the parameters $\{\alpha_i\}$ of (\ref{eq:complex_jastrow}) are determined by finding the ground state of a classical
2D $XY$ spin model. Once our wave function is optimized, we measure the physical observables, normalized by the number of copper atoms in
the lattice. To benchmark our wave function, we have compared its properties with those obtained by exact diagonalizations for $10$ holes on a small $8$ copper lattice with periodic boundary conditions (25\% hole doping), with very encouraging results (see \tref{benchmark1}).
Details will be presented elsewhere \cite{weber_current_variational_long}.
\begin{table*}
\begin{center}
\begin{tabular}{|c|c|c|c|c|c|c|c|c|}
%  \hline
  \hline
  wave function & $E_{tot}$ & $T_{dp}$ & $T_{pp}$ & $U_d$ & $U_p$ & $\Delta_p$ & $V_{dp}$ & variance   \\
  \hline
Lanczos     & $-1.13821$     & $-3.10036$  & $-0.79666$  & $0.26737$  & $0.08398$ & $1.77545$ & $0.63201$  & 0 \\
Jastrow wf  & $-1.0775(1)$   & $-3.06068$  & $-0.83073$  & $0.26176$  & $0.08197$ & $1.83466$ & $0.64076$ & $0.018$ \\
1 Lanczos step  & $-1.1153(1)$   & $-3.14070$  & $-0.83715$  & $0.26559$  & $0.08736$ & $1.86495$ & $0.64469$ & $0.018$ \\
Fixe node/ Jastrow  & $-1.1112(5)$   &             &             & $0.26966$  & $0.08708$ & $1.77314$ & $0.63177$ & $0.001$ \\
  \hline
%  \hline
\end{tabular}
\caption{Variational energies of the different wavefunctions compared with exact diagonalizations (Lanczos) on an $8$-copper lattice with
$10$ holes and $S^z=0$. We show the total energy ($E_{tot}$), the kinetic energy of the copper-oxygen links ($T_{dp}$) and of the oxygen-oxygen 
links ($T_{pp}$), the on-site repulsion energy of the d ($U_d$) and p ($U_p$) orbitals, the expectation value of the charge gap operator ($\Delta_p$), 
and the expectation value of the Coulomb repulsion between the $d$ and $p$ orbitals ($V_{dp}$). The results obtained by applying one Lanczos step 
(the best variational ansatz) and the fixed node approximation \cite{sorella_fixe_node} on a simple Jastrow wavefunction are also shown (the calculation of the off-diagonal operators $T_{dp}$ and $T_{pp}$ within the fixe-node approximation is more involved and is beyond the scope of the present study). The comparison shows that our
wavefunction captures quite well the low energy physics of the model.
}
\label{benchmark1}
\end{center}
\end{table*}

We now turn to larger lattices, using open boundary conditions to avoid the frustration of the $\alpha_i$ variables.
We have performed calculations on lattices ranging from $16$ to $64$ copper sites. The gain in energy for the best antiferromagnetic (SDW) instability and for flux phases are compared in \fref{fig:energies} \footnote{Remarkably, we found that
a wavefunction with BCS copper-oxygen and oxygen-oxygen pairing is stabilized close to $x=0.2$. The symmetry and the range
of stabilization will be discussed elsewhere \cite{weber_current_variational_long}.}
Note that these orders are somewhat exclusive: considering both simultaneously does not lead to any detectable
gain in energy.
At zero doping we find that the N\'eel magnetic long-range order is stabilized.
By introducing doping, we considered as a first approximation only the $Q=(\pi,\pi)$ pitch vector
for the spin density wave, even though the pitch vector is most likely doping dependent \cite{assad_spiral_3band}
or other instabilities like stripes can occur \cite{yanagisawa_vmc_3band}.
Nevertheless, the long-range correlations contained in the Jastrow factor allow for a decent treatment of the spin
correlations. Indeed for our best variational wave function, the magnetic order parameter $M=\lim_{r \rightarrow \infty}\sqrt{\langle S^z_i S^z_{i+r}\rangle})$ is  $\approx 60\%$ of the classical value. Using this wavefunction as a guiding function for the fixed node calculations, we find a slightly higher magnetic order ($62\%$). These values compare well with the $60\%$ obtained by quantum Monte Carlo in the one-band Heisenberg model \cite{Sandvik_2d_HB}. Note that, in the three-band Hubbard model, the magnetic instability is strongly dependent on the oxygen-oxygen hopping integral $t_{pp}$. Despite the good agreement for the order parameter, our variational treatment
overestimates the magnetic instability, which only disappears for $13\%$ doping, instead of the
experimentally observed $x=2\%$ for the cuprates. VMC tends indeed to overestimate the stability of SDW phases because the alternating magnetization allows to avoid double occupancy in the uncorrelated part of the wavefunction. The presence of magnetic order clearly costs kinetic energy, but it does a better job than a pure local Gutzwiller projection, which reduces the kinetic energy much more dramatically.

Let us now turn to the current instability. On intermediate system sizes such as the $16$ copper lattice,
we find that an orbital current phase with a symmetry $\Theta_2$ (see the inset of \fref{fig:energies}) is stabilized. Our wave function thus shows a tendency to flux phase that does not exist in the corresponding one band model. However the gain in energy strongly decreases as the size increases and seems, within the accuracy of our calculation,
to vanish in the thermodynamic limit. Taken literally, this suggests that the orbital current phase is not stable, and that the system only has short range correlations. However we of course cannot rule out that some fine tuning of the parameters could stabilize this phase, or that the energy gain would be much smaller than our statistical error. Two points should be emphasized. First, regardless of the size of the system, we find consistently that the $\Theta_2$ symmetry is the one with the lowest variational energy. Other symmetries such as the $\Theta_1$ phase or the DDW patterns are unstable within our variational approach in the whole range of doping. Second, varying the Cu-O interaction seems to have no major effect on the stability of the current patterns, in contrast with what one could have
expected on the basis of the mean-field solution \cite{varma_first_time_mf_theta2}.
Our study demonstrates that for an instability towards long-range current correlations to develop, the relative sign of the transfer integrals around a loop is a crucial parameter. Indeed, when the
sign of $t_{pp}$ is reversed, $J$ and $J_{MF}$ are oriented in the same direction, and the current pattern is stable.
In the mean field solution, this change of sign is induced by the correction to the bare $t_{pd}$ coming from the decoupling of the Cu-O repulsion term.  

Given our variational results for the three-band Hubbard model of CuO$_2$ planes, and the experimental evidence, it is natural to check if other terms not included into this simple version of a multi-band 
model could produce such an effect and stabilize orbital currents.
One obvious possibility is to include correlated hopping, as already done successfully for the single-band Hubbard model. This indeed leads to a strong increase of the tendency to develop orbital currents \cite{weber_current_variational_long}.
Note however that the magnitude of these terms is difficult to assess.
\begin{figure}
  \begin{center}
    \center{\includegraphics[width=\miniwidth]{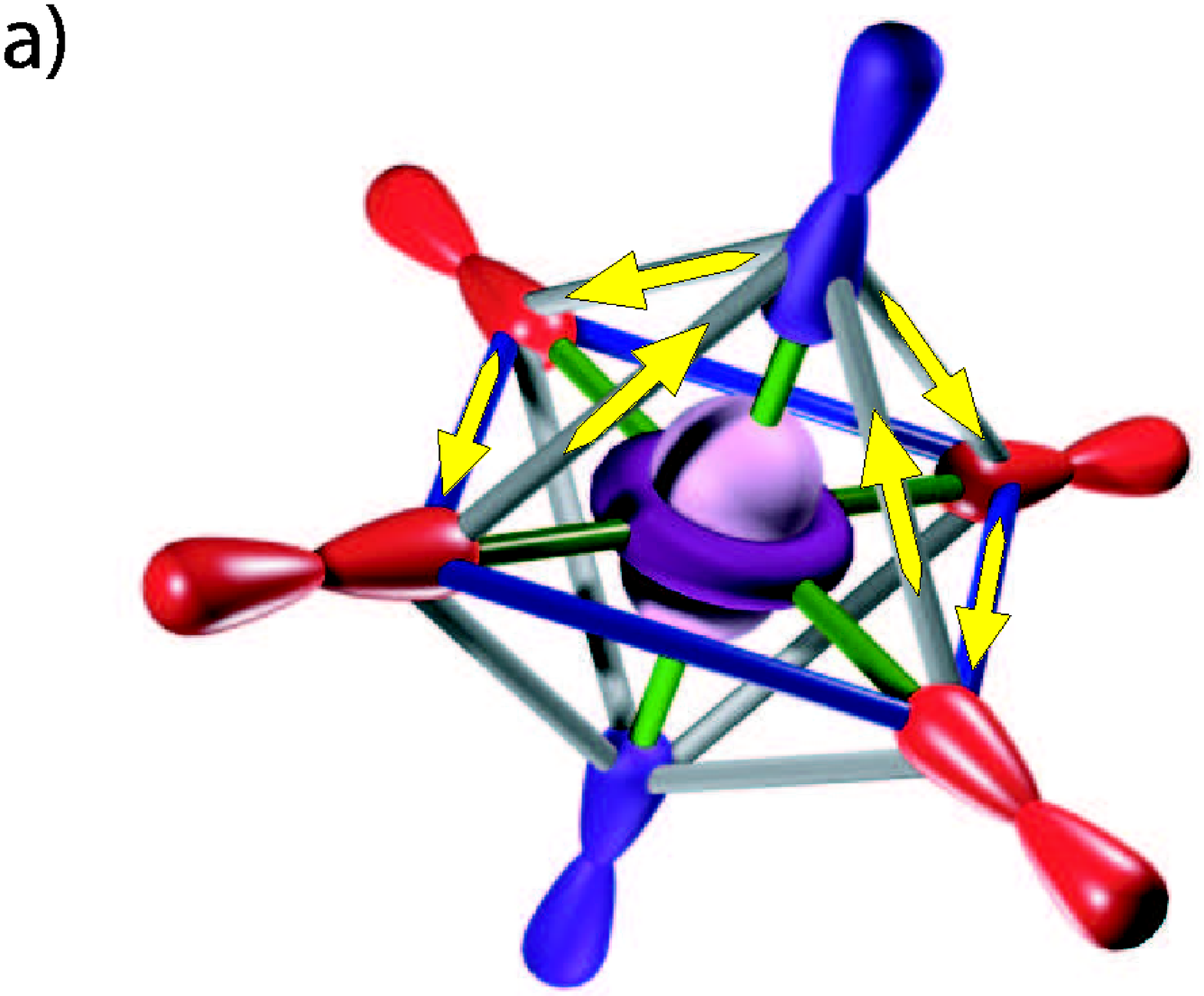}
    \includegraphics[width=\miniwidth]{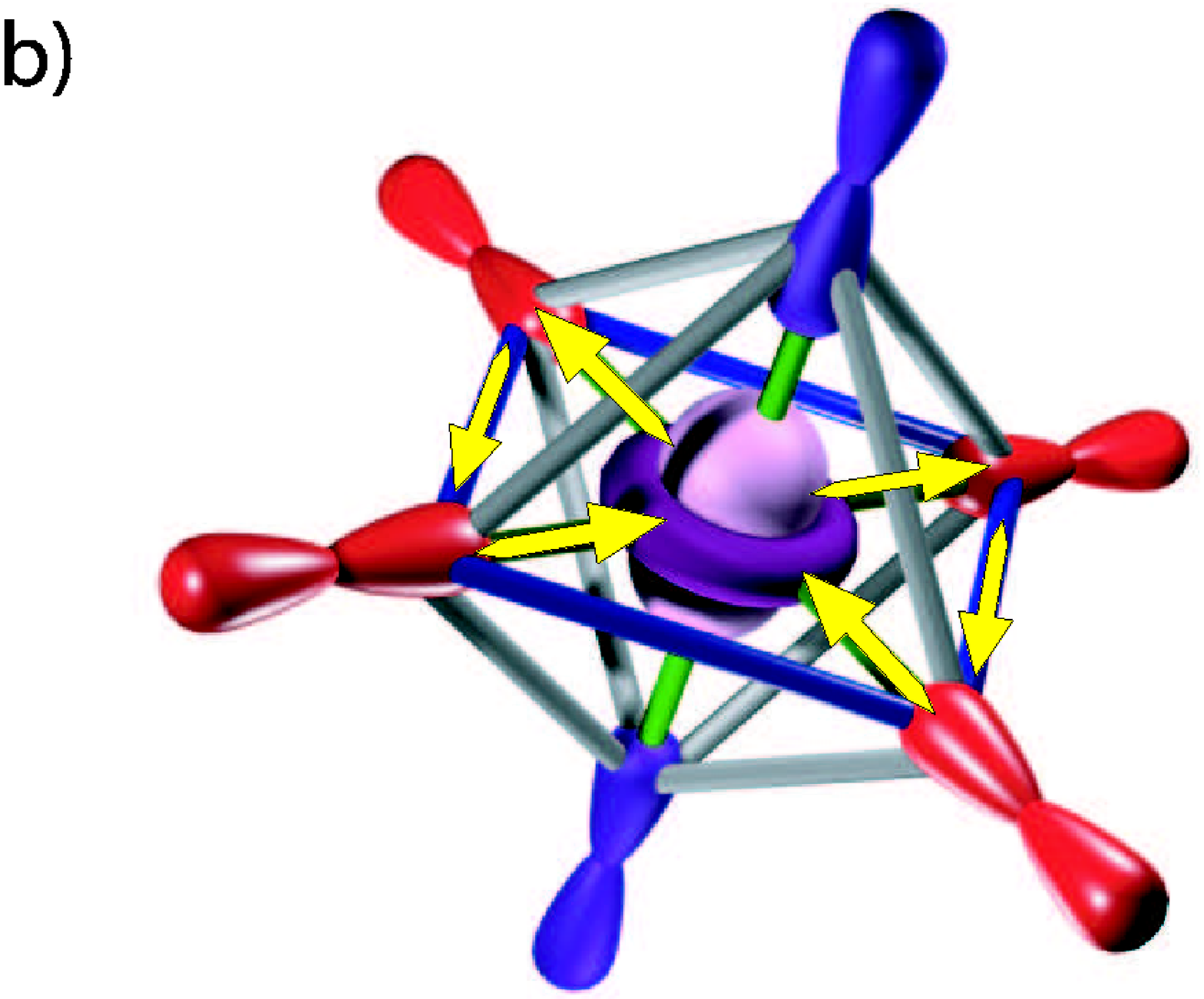}}
    \includegraphics[width=\normwidth]{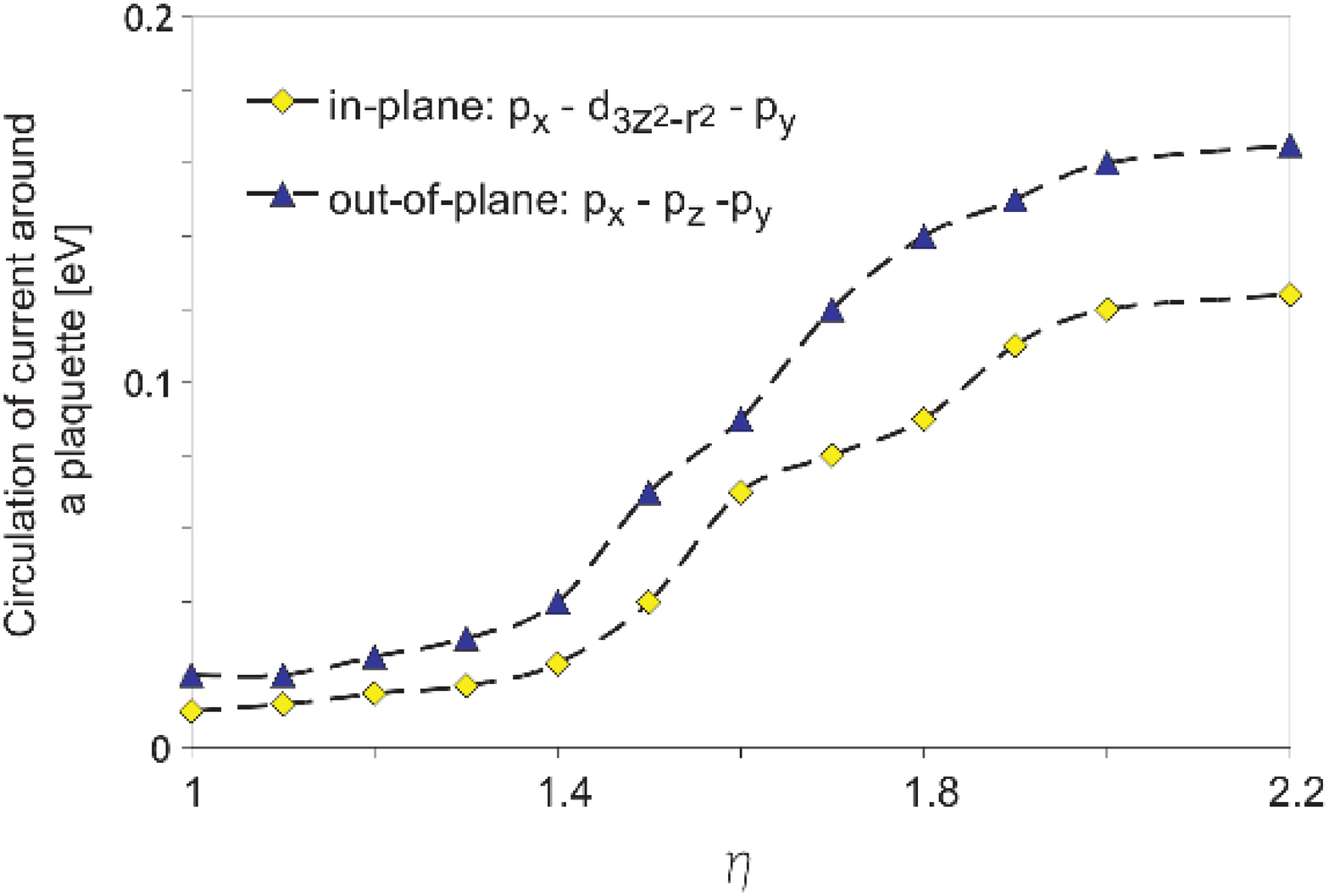}
    \caption{Circulation of current around a $p_x - p_z - p_y$ plaquette (triangles) and around
    a $p_x - d_{3z2-r2} - p_y$ plaquette (squares) measured in our best variational Ansatz for the eight band Hubbard model.
    The phenomenological parameter $\eta$ renormalizes the amplitudes of the out-of-plane transfer integrals.
   The symmetry of the orbital current pattern is $\Theta_2$ like: there are two out-of-plane current loops in the upper pyramid (a) and two current loops in the copper plane (b). Finally, the current pattern in the lower pyramid (not shown) is obtained by a horizontal mirror reflection of the upper pyramid. The calculations were done on a 32 Cu lattice at $x=0.125$ hole doping.}
 \label{fig:apical}
  \end{center}
\end{figure}
A second and interesting possibility is to generalize the model by including apical oxygens. Indeed, the relative signs of the hopping around purely oxygen plaquettes allow \emph{a priori} for orbital currents. We have thus repeated the calculation for Cu-O layers including apical oxygens above and below each Cu atom as well as the Cu-\dz orbitals. The bare parameters have been taken from
LDA \cite{mcmahancuprate} performed with the crystal structure of the insulating parent compound, and the analysis has been carried
out as a function of the distance between the copper and the apical oxygen, modelled by a renormalization
parameter $\eta$ that multiplies the hopping integrals involving the apical oxygens. As expected from the signs of
the hopping, a calculation on a 32-Cu lattice (192 sites) with $x=0.125$ hole doping confirms that orbital currents involving the apical oxygen or the Cu-\dz indeed develop according to the pattern of  \fref{fig:apical}. These currents are quite small for the bare values of the parameters, but they strongly increase to reach significant values above $\eta=1.5$. Interestingly enough,
the current circulating in the $p_x - p_z - p_y$ plaquette leads to a tilted moment, which would provide a natural explanation for the out of plane moment that was observed in the neutron experiments \cite{fauque_neutrons_currents}. Whether the structural changes induced by doping on the position of the apical oxygens reported by some authors can be large enough to produce such a strong renormalization of the hopping integrals remains to be seen \cite{jahn_teller_cuprates}.

\begin{acknowledgements}
We thank C. M. Varma and P. Bourges for many illuminating discussions. One of us (CW) is also grateful to G. Kotliar for enlightening discussion. This work was supported in part by the Swiss NSF under MaNEP and Division II.
\end{acknowledgements}

\bibliographystyle{prsty}

\begin{thebibliography}{10}

\bibitem{bonn_hightc_review}
D. Bonn, Nature Phys. {\bf 2},  159  (2006).

\bibitem{affleck_marston}
I. Affleck and J.~B. Marston, Phys. Rev. B {\bf 37},  3774  (1988).

\bibitem{coleman_flux_2band}
J. Mydosh and P. Chandra and P. Coleman and V. Tripathi, Acta Physica Polonica B, 34, 659-665 (2003).

\bibitem{coleman_flux_2bandb}
P. Chandra and P. Coleman and J. A. Mydosh and V. Tripathi, Nature, 417, 831-834 (2002). 

\bibitem{kotliar_liu_dwave_slavebosons}
G. Kotliar and J. Liu, Phys. Rev. B {\bf 38},  5142(R)  (1988).

\bibitem{chakravarty_ddw_pseudogap}
S. Chakravarty, R.~B. Laughlin, D.~K. Morr, and C. Nayak, Phys. Rev. B {\bf
  63},  094503  (2001).

\bibitem{varma_first_time_mf_theta2}
M.~E. Simon and C.~M. Varma, Phys. Rev. Lett. {\bf 89}, 247003 (2002)

%\bibitem{varma_meanfield_currents_pseudogap}
%C.~M. Varma, Phys. Rev. B {\bf 73},  155113  (2006).

%\bibitem{varma_3band_currents}
%C.~M. Varma, Phys. Rev. B {\bf 55},  14554  (1997).

\bibitem{varma_pseudogap_theory}
C.~M. Varma, Phys. Rev. Lett. {\bf 83}, 3538 (1999)

\bibitem{fauque_neutrons_currents}
B. Fauqu{\'e} {\it et~al.}, Phys. Rev. Lett. {\bf 96},  197001  (2006).

\bibitem{xia_polarkerr_pseudogap}
J. Xia {\it et~al.}, cond-mat:arXiv  0711.2494  (2007).

\bibitem{laderersuperconductivity_flux_ref}
P. Lederer, D. Poilblanc, and T.~M. Rice, Phys. Rev. Lett. {\bf 63},  1519
  (1989).

\bibitem{zhang_flux_ref}
F.~C. Zhang, Phys. Rev. Lett. {\bf 64},  974  (1990).

\bibitem{orignac_2chain_long}
E. Orignac and T. Giamarchi, Phys. Rev. B {\bf 56},  7167  (1997).

\bibitem{schollwock_ladder_strong_fluxphases}
U. Schollwock {\it et~al.}, Phys. Rev. Lett. {\bf 90},  186401  (2003).

\bibitem{sudbo_cuo}
A. Sudb{\o}\ {\it et~al.}, Phys. Rev. Lett. {\bf 70},  978  (1993).

\bibitem{lee_marston_CuO}
S. Lee, J.~B. Marston, and J.O. Fjaerestad, Phys. Rev. B {\bf 72},  075126
  (2005).

\bibitem{chudzinski_ladder_rapid}
P. Chudzinski, M. Gabay, and T. Giamarchi, Phys. Rev. B {\bf 76},  161101(R)
  (2007).

\bibitem{greiter_exactdiag_currents}
M. Greiter and R. Thomale, Phys. Rev. Lett. {\bf 99},  027005  (2007).

\bibitem{greiter_exactdiag_currents2}
M. Greiter and R. Thomale, Phys. Rev. B {\bf 77},  094511  (2008).

\bibitem{hybertsen_DFT_parameters_cuprates}
M.~S. Hybertsen, E.~B. Stechel, M. Schluter, and D.~R. Jennison, Phys. Rev. B
  {\bf 41},  11068  (1990).

\bibitem{emery_3bands_model}
V.~J. Emery, Phys. Rev. Lett. {\bf 58},  2794  (1987).

\bibitem{mila_parameters_cuprates}
F. Mila, Phys. Rev. B {\bf 38},  11358  (1988).

\bibitem{yanagisawa_vmc_3band}
T. Yanagisawa, S. Koike, and K. Yamaji, Phys. Rev. B {\bf 64},  184509  (2001).

\bibitem{oguri_variational_3band}
A. Oguri, T. Asahata, and S. Mackawa, Physica B {\bf 186-188},  953  (1993).

\bibitem{Trivedi_complex_Jastrow}
S. Liang, N. Trivedi, Phys. Rev. Lett. {\bf 64},  232  (1990).

\bibitem{umrigar_mcv_minimis}
C.~J. Umrigar, K.~G. Wilson, and J.~W. Wilkins, Phys. Rev. Lett. {\bf 60},
  1719  (1988).

\bibitem{sorella_optimization_VMC}
S. Sorella, Phys. Rev. B {\bf 71},  241103(R)  (2005).

\bibitem{weber_current_variational_long}
C. Weber, A. Laeuchli, F. Mila, and T. Giamarchi, 2008, in preparation.

\bibitem{sorella_fixe_node}
S. Sorella and L. Capriotti, Phys. Rev. B {\bf 61},  2599  (2000).

\bibitem{assad_spiral_3band}
F.~F. Assaad, Phys. Rev. B {\bf 47},  7910  (1993).

\bibitem{Sandvik_2d_HB}
A.~W. Sandvik, Physical Review B (Condensed Matter) {\bf 56},  11678  (1997).

\bibitem{mcmahancuprate}
A.~K. McMahan, J. f.~Annett, and R.~M. Martin, Phys. Rev. B {\bf 42},  6268
  (1990).

\bibitem{jahn_teller_cuprates}
H. Kamimura, H. Ushio, and S. Matsuno, Springer-Verlag Berlin Heidelberg  157
  (2007).

\end{thebibliography}

\end{document}